# Spatial mapping and analysis of aerosols during a forest fire using computational mobile microscopy

Yichen Wu[1,2,3,*], Ashutosh Shiledar[1], Yi Luo[1,2,3], Jeffrey Wong[4], Cheng Chen[1], Bijie Bai[1], Yibo Zhang[1,2,3], Miu Tamamitsu[1,2,3], and Aydogan Ozcan[1,2,3,5,§]

[1] Electrical Engineering Department, University of California, Los Angeles, CA, 90095, USA.
[2] Bioengineering Department, University of California, Los Angeles, CA, 90095, USA.
[3] California NanoSystems Institute (CNSI), University of California, Los Angeles, CA, 90095, USA.
[4] Computer Science Department, University of California, Los Angeles, CA, 90095, USA.
[5] David Geffen School of Medicine, University of California, Los Angeles, CA, 90095, USA

[*] wuyichen@ucla.edu  [§] ozcan@ucla.edu  http://innovate.ee.ucla.edu/

## ABSTRACT

Forest fires are a major source of particulate matter (PM) air pollution on a global scale. The composition and impact of PM are typically studied using only laboratory instruments and extrapolated to real fire events owing to a lack of analytical techniques suitable for field-settings. To address this and similar field test challenges, we developed a mobile-microscopy- and machine-learning-based air quality monitoring platform called c-Air, which can perform air sampling and microscopic analysis of aerosols in an integrated portable device. We tested its performance for PM sizing and morphological analysis during a recent forest fire event in La Tuna Canyon Park by spatially mapping the PM. The result shows that with decreasing distance to the fire site, the PM concentration increases dramatically, especially for particles smaller than 2 μm. Image analysis from the c-Air portable device also shows that the increased PM is comparatively strongly absorbing and asymmetric, with an aspect ratio of 0.5–0.7. These PM features indicate that a major portion of the PM may be open-flame-combustion-generated element carbon soot-type particles. This initial small-scale experiment shows that c-Air has some potential for forest fire monitoring.

**KEYWORDS:** aerosol, black carbon, forest fire, c-Air, computational microscopy, digital holography

# 1. INTRODUCTION

Forest fires are a significant source of particulate matter (PM) air pollution on a global scale.[1] PM from forest fires can be transported over long and even continental distances. This PM can negatively affect the environment, including climate change;[2] it can cause various respiratory and cardiovascular diseases in humans and is also a carcinogen.[3]

Combustion-generated PM from forest fires has a unique chemical composition and morphology, and is typically smaller than geologically produced dust.[1] The physical and chemical properties of combustion-generated PM have been studied extensively in laboratory settings,[1–13] where wood and other types of biofuels are burned in a chamber[2,5,7,9,10] or a wood stove,[2,4,6] and the resulting particles are sampled using filtering and/or an impactor and then analyzed using electron microscopy to study their size and morphology[2,5,7,9,14] and using spectroscopy and thermal-optical techniques to study their chemical composition.[2,4,7,9,12,14] The nature of combustion-generated PM depends heavily on the fuel type,[5,7,8] as well as the combustion conditions such as temperature.[4,5,9,14] Specifically, dry fuels burning at high temperature tend to generate chain-like agglomerates of spherical inorganic carbon particles, which are known as soot.[2,4,5,7,11] Soot features high optical absorption,[6,9] irregular shapes, and a fractal structure.[2,3,5,7,14] In contrast, wet fuels burning at lower temperature tend to generate condensed organic carbon balls from volatile matter, which are sometimes called tar balls.[2,4–6,11] Tar balls typically exhibit low optical absorption[6,9] and are semiliquid and spherical.[5,6,11,14] In a real forest fire, there is usually a mixture of both combustion conditions,[9] and the different aerosols can be viewed as fingerprints to track these conditions.

Unlike that in well-controlled laboratory experiments, emission from real wildfires is poorly characterized owing to the variability in combustion conditions, e.g., wind propagation, fuel loading, and fuel moisture. As a result, the PM from a forest fire and its impact are poorly understood. The impact of a wildfire is usually estimated by simply multiplying the total fuel load at the fire site by the laboratory-measured emission factor of each fuel.[1] This estimate is obviously very rough because it is quite questionable whether a laboratory experimental result can scale up to a real forest fire. Moreover, this question remains unverified owing to a lack of portable devices that can characterize these aerosols in real time during a fire event.

As a disruptive technology, we developed a mobile-microscopy- and machine-learning-based portable PM monitor, c-Air.[15,16] It is based on an impactor to capture aerosols and a computational lensless microscope[17–30] to image and analyze their microscopic features in real time. It dynamically samples the air at 13 L/min and performs sizing and microscopic analysis of aerosols down to ~1 µm using digital holographic reconstruction and machine learning with high accuracy and high throughput.[15]

Recently, a forest fire struck La Tuna Canyon Park in the greater Los Angeles area, California, and was later known as the La Tuna Fire.[31,32] As this forest fire progressed, we used our c-Air device and took a series of measurements at different geographical locations. The image processing result of the c-Air measurements shows that a significant portion of the aerosols that the La Tuna Fire generated have comparatively small size, high absorption, and noncircular shapes, which suggests that they are irregularly shaped soot-type black carbon from a hot-burning and flaming event.[1,2,4,5,8]

Because c-Air is portable and computationally enhanced, we envision it playing a major role in future monitoring of forest fire events to detect and quantify different forest fire trace particles and analyze their impact in a way that is conventionally not accessible.

# 2. MATERIALS AND METHODS

## 2.1 c-Air: lensless-microscope-based aerosol monitor

c-Air is a computational platform that combines an impactor air sampler (Air-O-Cell, Zefon International) with a lensless microscope, as shown in Fig. 1. We refer to the basic imaging geometry as impactor-on-a-chip because the impactor nozzle, together with its sticky collection coverslip, is placed directly on top of a complementary metal-oxide semiconductor (CMOS) image sensor (OV5647, Omnivision, 5 MP pixels), with a spacing of approximately 400 µm between the nozzle and sensor. An air pump (GTEK M00198, 13 L/min) drives air through the impactor nozzle, and aerosols within the air stream strike and are collected on the top of the sticky coverslip. On the top of the device, three fiber-coupled light-emitting diodes (LEDs) illuminate the sticky coverslip, and the captured aerosols cast an in-line hologram, which is recorded by the CMOS image sensor chip. These holograms are used for reconstruction[33–35] and analysis of the aerosols, as described in the following subsections. The device is powered by a rechargeable lithium polymer battery and controlled and automated by a microprocessor (Raspberry Pi A+).

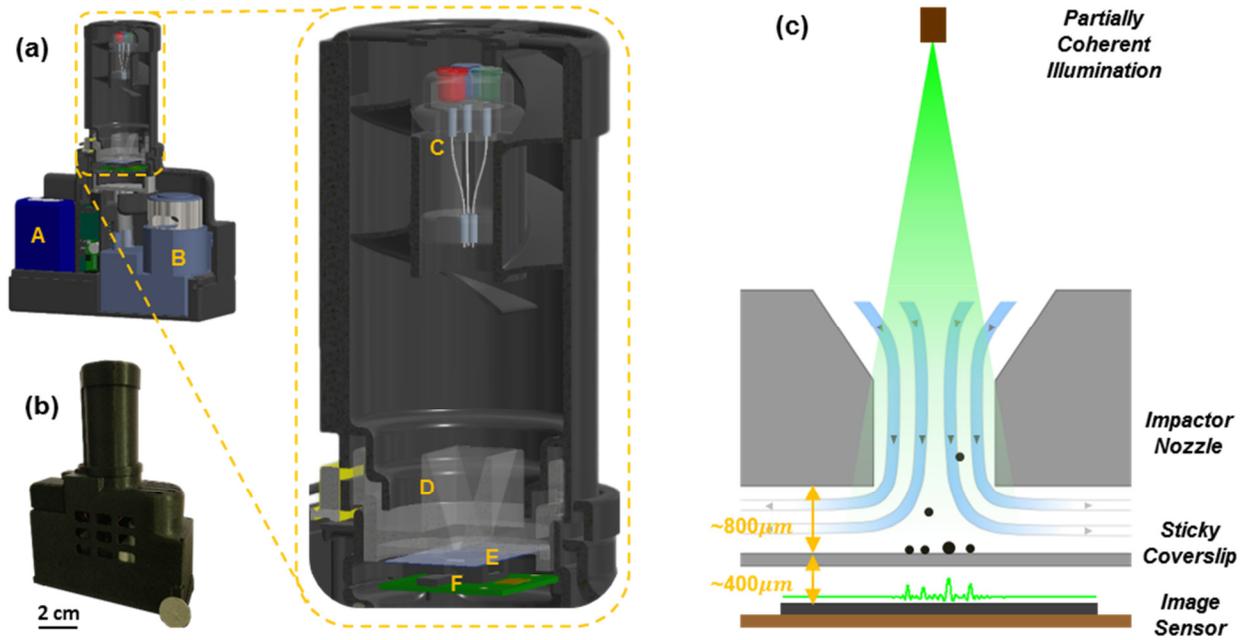

**Fig. 1. Lens-free-microscope-based air-sampler, c-Air.** (a) 3D computer-aided-design (CAD) overview of the device, including (A) rechargeable battery, (B) vacuum pump (13 L/min), (C) illumination module with fiber-coupled LEDs in red (624 nm), green (527 nm), and blue (470 nm), (D) impaction-based air sampler with (E) a sticky coverslip on top of (F) the image sensor. (b) Photo of the c-Air device. A US 25-cent coin (quarter) is placed next to the device for scale. (c) Schematic drawing of impaction-based air sampler on a chip. The device is powered by a rechargeable battery (A) and controlled by a microcontroller (Raspberry Pi A+). The assembly and packaging are 3D-printed.

**2.2 Particle sizing using a commercial optical particle counter (OPC) and c-Air**

Particle sizing is performed at three spatial locations, namely, location 1 (9042 Wildwood Avenue), location 2 (9760 La Tuna Canyon Road), and location 3 (9078 La Tuna Canyon Road), as shown in Fig. 2(a), using two platforms: a commercial optical particle counter (OPC) and c-Air. The OPC samples the air through a small channel at 2.83 L/min. A laser beam is focused on the aerosols passing through the channel and is scattered by these aerosols. A photodetector records the scattering intensity, which is used to infer the particle sizes on the basis of their scattering cross section. The aerosol counts are grouped into four size categories: 0.5–1.0 μm, 1.0–2.5 μm, 2.5–5 μm, and >5 μm.

The c-Air device samples the air through an impactor and performs microscopy and image analysis of the captured particles on the impactor. A peeling algorithm and trained machine learning algorithms are used to detect and predict a size for each individual particle, with a sizing accuracy of ~93%. (See the methods section in Ref. [15] for more details.) The particle sizes are then plotted as a histogram from 1 to 20 μm, with a 2 μm spacing.

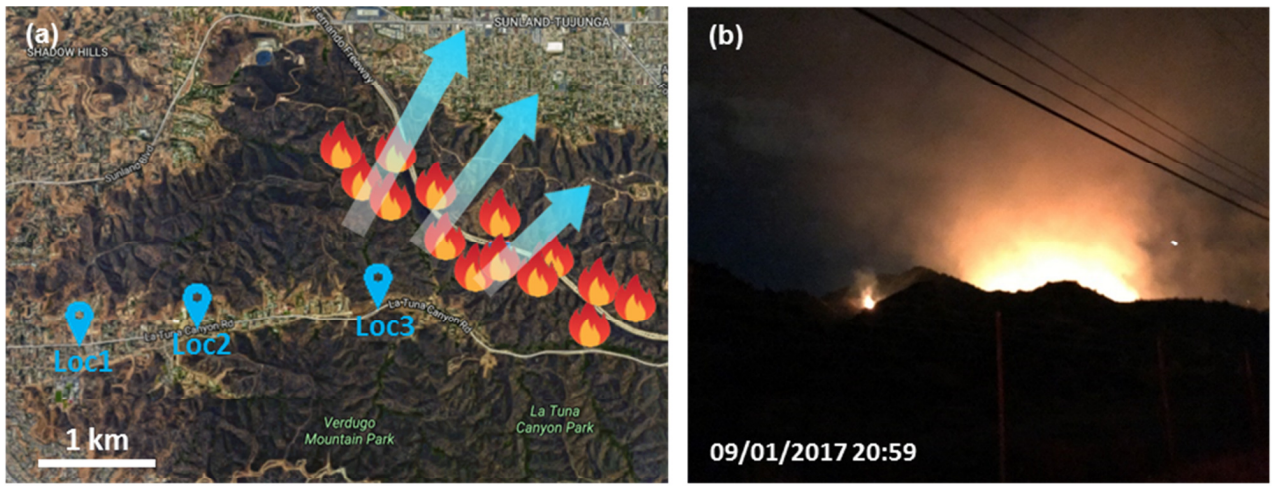

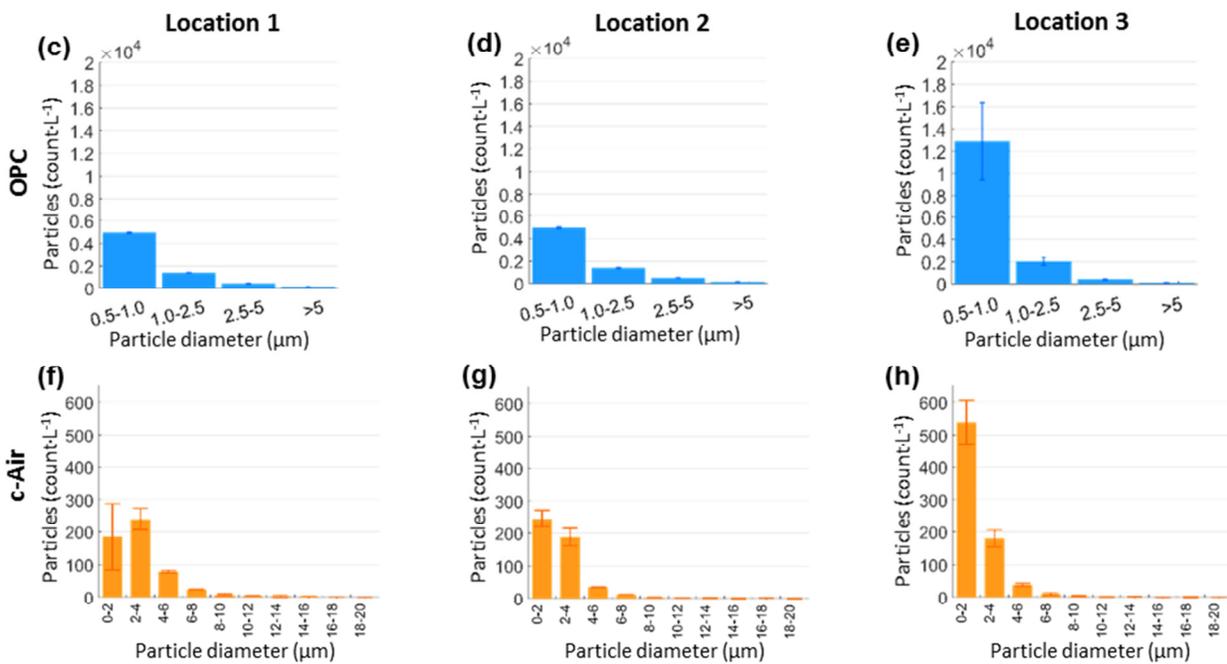

**Fig. 2. Particle sizing results at different locations.** (a) Map of measurement locations during the La Tuna Fire on 09/01/2017. The blue arrows mark the wind direction. (b) Photograph of forest fire on top of a mountain that was witnessed remotely at measurement location 3. (c–e) Particle sizing results at locations 1–3 using a commercial OPC. (f–h) Particle sizing results at the same locations using c-Air device. A significant increase is observed, especially in smaller particles, with decreasing distance to the fire site.

### 2.3 Particle feature analysis based on c-Air images

Automatic morphological analysis is performed for each reconstructed and detected particle. An intensity level (0.95) and a phase level (0.12 rad) are defined as thresholds for these particles, which are three times the standard deviation away from the mean of the background. If a particle has an intensity of less than 0.95, it is marked as an amplitude particle, and an amplitude mask is calculated. If the particle has a phase level greater than 0.12 rad, it is marked as a phase particle, and a phase mask is calculated.

Because most of the particles (>99%) are either amplitude-only or have a strong amplitude signature [as shown in Fig. 3(a)], we use the amplitude mask to perform morphological analysis using the MATLAB function "regionprops." Two major characteristics are analyzed: the threshold area and aspect ratio (AR). The threshold area is defined as the total number of pixels times the metric area of each pixel. The aspect ratio is defined as the ratio of the length of the shorter axis ($L_{min}$) to that of the longer axis ($L_{max}$) of the threshold region, i.e., AR = $L_{min}/L_{max}$.

## 3. RESULTS AND DISCUSSION

On 09/01/2017, during the La Tuna forest fire,[31(p500),32] we drove to three locations and took aerosol samples there; these are location 1 (9042 Wildwood Avenue), location 2 (9760 La Tuna Canyon Road), and location 3 (9078 La Tuna Canyon Road), as illustrated in Fig. 2(a). At locations 1 and 2, we took three 30 s samples of air using the c-Air device at a flow rate of 13 L/min. At location 3, we took six 30 s samples of air at a flow rate of 13 L/min. The sample images were then sent to our remote server and analyzed as described in the methods section. A commercial OPC (Met One 804) was also used to count the particles in four different size channels: 0.5–1.0 μm, 1.0–2.5 μm, 2.5–5 μm, and >5 μm. During the measurement, the wind was blowing toward the northeast. It is reasonable to believe that there were much higher concentrations of PM on the downwind side of the fire. Unfortunately, because the road was blocked, we could not take further measurements closer to the fire site or at downwind locations.

Fig. 2 plots the particle sizing results. From Fig. 2(f–h), we see that with decreasing distance to the fire site (especially at location 3), the total sampled PM concentration increases, and most of the particles contributing to this increase are particles smaller than 2 μm, the count density of which is more than doubled. A similar trend is also observed by the commercial OPC, which shows an increase in particles with sizes of 1.0–2.5 μm and a significant increase in particles with sizes of 0.5–1.0 μm [as shown in Fig. 2(c-e)]. This is consistent with previously reported laboratory test results, where open-flame-generated black-carbon agglomerates (soot) were usually smaller than geologically produced dust,[1,8] with a typical size range of 20–2000 nm.[3,5]

Fig. 3(a) shows pie charts of the particle composition percentages at each location according to the amplitude and phase signature strength, as described in the methods section. As shown in the examples in Fig. 3(b), the phase particles usually exhibit less or nearly no absorption of illumination light but leave a phase signature, as their refractive index differs from that of the ambient media. These phase particles can be bio-aerosols such as some pollens or bacteria, or in this case organic carbon from the smoldering phase.[6,9] In contrast, amplitude particles are those with strong absorption of illumination light, such as very dark elemental carbon generated by open-flame fires.[6,9] Comparing the pie charts in Fig. 3(a), we see that with decreasing distance to the fire site, the portion of amplitude particles increases, indicating that the additional aerosols are strongly absorbing amplitude particles, which is another indication of soot-type particles generated by open-flame fires.

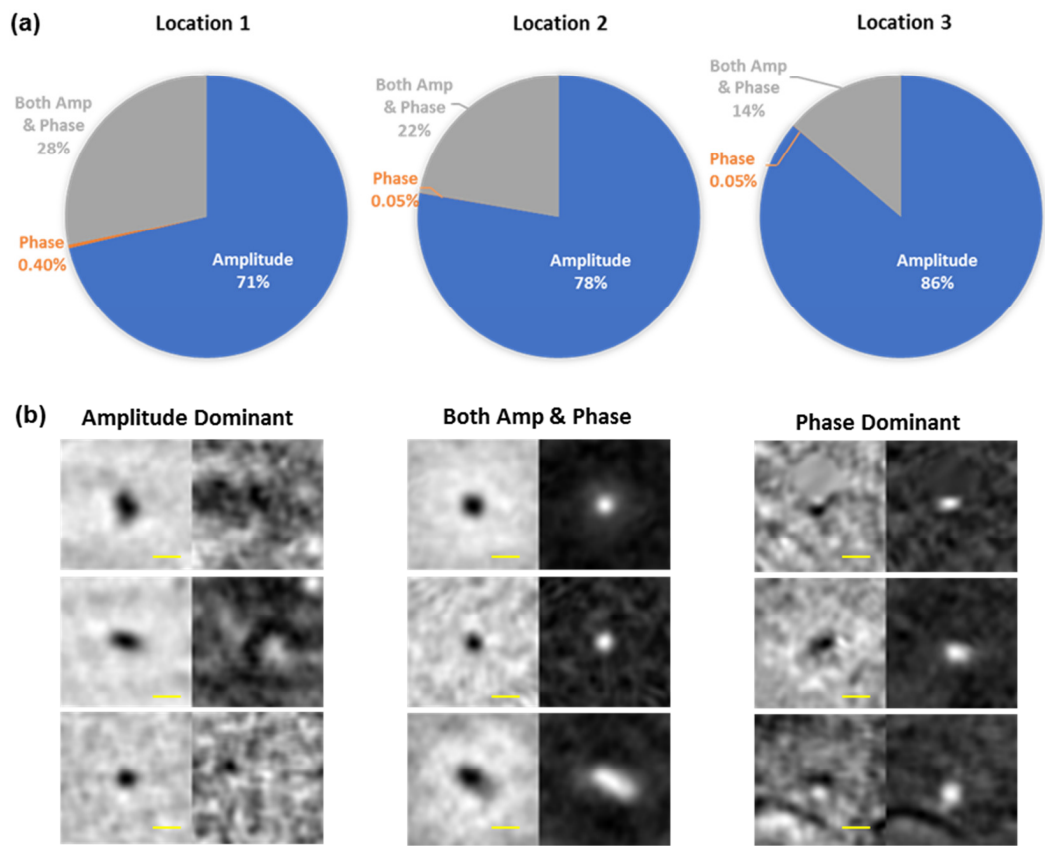

**Fig. 3. Particle feature classification.** (a) Pie-charts of PM classification at different spatial locations. The aerosols sampled at locations 1–3 are classified as amplitude particles, phase particles, and both amplitude and phase particles according to their amplitude and phase signal strength. With decreasing distance to the fire site, the proportion of amplitude particles increases, showing strong absorption by these fire-generated aerosols. (b) Several examples of reconstructed images of these three classes of particles. For each particle, the left image is an amplitude reconstruction, whereas the right image is a phase reconstruction. All the amplitude and phase images shown here are normalized. Scale bars: 5 μm.

We further plot the morphological analysis result of the sampled particles in Fig. 4. Two morphological features are selected: the threshold area and aspect ratio. The threshold area is related to the size of a particle, whereas the aspect ratio reveals the shape and asymmetry of a particle. The top row of Fig. 4 shows that with decreasing distance to the fire site, the smaller-area aerosols increase, which is consistent with the PM sizing results. The bottom row of Fig. 4 shows that the increase in PM concentration with decreasing distance to the fire site is caused by particles that are neither circular nor symmetric. Instead, they have an aspect ratio of approximately 0.5–0.7. The asymmetric nature of the particles is a significant indicator of irregular fractal-shaped soot particles, in contrast to circular organic carbon.[2,5,8] More specifically, it has also been reported in a laboratory combustion experiment of wildland fuels that the generated soot particles have an aspect ratio of 0.49–0.63, in agreement with our field-test results.[5] In addition, we note that this asymmetric nature of PM is usually ignored by conventional OPCs, the sizing model of which is based on spherical particles. This may cause some errors in OPC results compared to the gold standard of microscopic analysis.

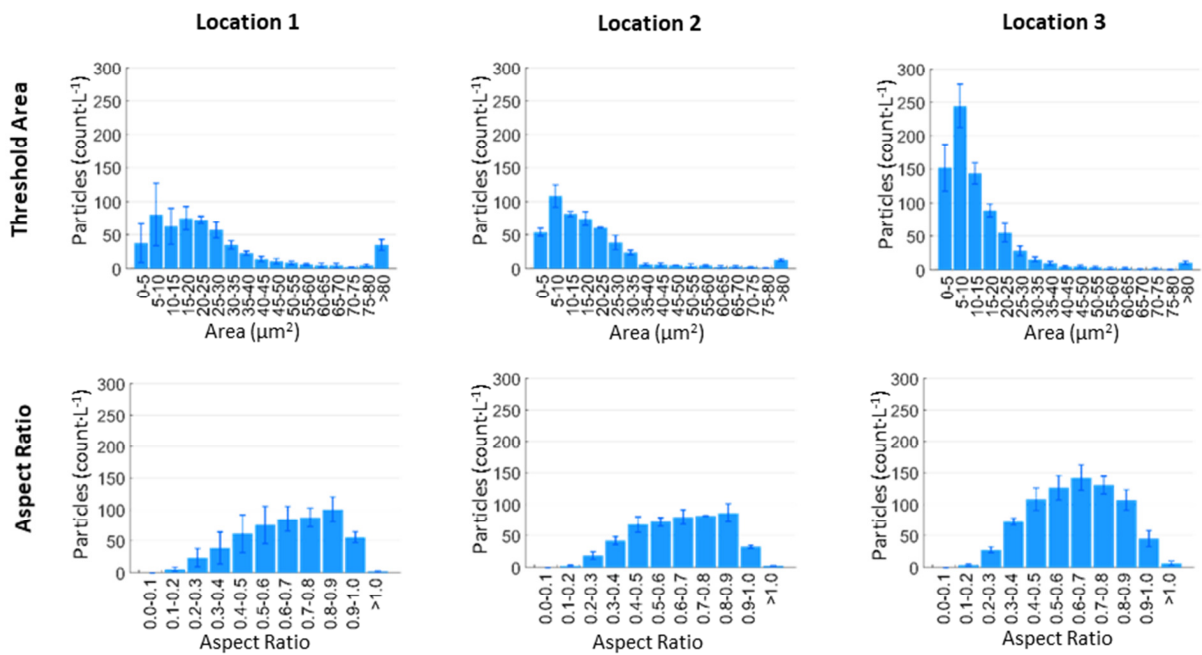

**Fig. 4. Particle morphological analysis by c-Air.** (Top row) Threshold area of PM samples captured at locations 1–3. (Bottom row) Aspect ratio of PM at the same locations. The increase with decreasing distance to the fire site consists mostly of particles with smaller areas and asymmetric shapes, with an aspect ratio of approximately 0.6.

## 4. CONCLUSION

In conclusion, we performed small-scale spatial mapping of PM during the recent La Tuna forest fire. Three sample locations upwind of the fire sites were selected, and the aerosols at these locations were measured by the portable computational c-Air device as well as a commercial OPC. The particle sizing result of c-Air agreed with that of the commercial OPC, showing more than twice as many smaller particles (<2 µm) at a location closer to the fire site compared to that at locations that were further away. Image analysis of captured aerosols showed that with decreasing distance to the fire site, the portion of strongly absorbing amplitude particles increased. It also showed that most of the additional particles were asymmetric ones with an aspect ratio of 0.5–0.7. The small particle size, high absorption, and asymmetric particle shape suggests that the increased PM concentration may be related to soot-type particles of elemental carbon generated during open-flame combustion in the forest fire, as suggested by other independent *laboratory* experiments. These features may be used as fingerprints to trace the presence and impact of forest-fire-generated PM. This study shows that the c-Air device has some potential for on-site forest fire monitoring.

## ACKNOWLEDGEMENTS

The authors acknowledge the support of the Presidential Early Career Award for Scientists and Engineers (PECASE), the Army Research Office (ARO; W911NF-13-1-0419 and W911NF-13-1-0197), the ARO Life Sciences Division, the National Science Foundation (NSF) CBET Division Biophotonics Program, the National Science Foundation (NSF) Emerging Frontiers in Research and Innovation (EFRI) Award, the NSF EAGER Award, the NSF INSPIRE Award, the NSF Partnerships for Innovation: Building Innovation Capacity (PFI:BIC) Program, the Office of Naval Research (ONR), the National Institutes of Health (NIH), the Howard Hughes Medical Institute (HHMI), the Vodafone Americas Foundation, the Mary Kay Foundation, the Steven & Alexandra Cohen Foundation, and KAUST. This work is based on research performed in a laboratory renovated by the NSF under Grant No. 0963183, which is an award funded under the American Recovery and Reinvestment Act of 2009 (ARRA). The authors also acknowledge Dr. Zoltán Göröcs from UCLA for providing information on the La Tuna Forest Fire.